\documentclass[12pt,letterpaper]{article}

\usepackage{amsmath}
\usepackage{amsfonts}
\usepackage{amssymb}
\usepackage{graphicx}

\renewcommand\[{\left[}

\newcommand\ee{\end{equation}}

\newcommand\be{\begin{equation}}

\newcommand\eea{\end{eqnarray}}

\newcommand\bea{\begin{eqnarray}}

\begin{document}

\vspace*{1cm}
\begin{flushright}
DFPD-08-A-09
\end{flushright}

\begin{center}

\def\thefootnote{\fnsymbol{footnote}}
{\Large \bf On the non-Gaussianity from Recombination}
\vskip 0.5cm
{\large Nicola Bartolo$^{\rm a,b}$ and Antonio Riotto$^{\rm b,c}$ \\[.1cm]}
\vskip 0.5cm

{\small \textit{$^{\rm a}$ }Dipartimento di Fisica ``G.\ Galilei'', Universit\`{a} di Padova, via Marzolo 8, Padova I-35131, Italy}

\vspace{.2cm}

{\small \textit{$^{\rm b}$ }INFN, Sezione di Padova, via Marzolo 8, Padova I-35131, Italy}

\vspace{.2cm}

{\small \textit{$^{\rm c}$ }CERN, PH-TH Division, CH-1211 Geneva 23, Switzerland}

\end{center}

\vspace{.8cm}

\hrule \vspace{0.3cm}
{\small  \noindent \textbf{Abstract} \\[0.3cm]
\noindent 
The non-linear effects operating at
the recombination epoch generate a non-Gaussian signal in the CMB anisotropies.  
Such a contribution is  relevant because it 
represents a major part of the second-order radiation transfer function which must be determined in order to have a 
complete control of both the primordial and non-primordial part of non-Gaussianity in the CMB anisotropies.
We provide an estimate of the level of non-Gaussianity in the CMB arising from the recombination epoch which   
shows up mainly in the equilateral configuration. We find that it causes a contamination to the possible measurement 
of the equilateral 
primordial bispectrum
shifting the minimum detectable value of the non-Gaussian parameter $f^{\rm equil}_{\rm NL}$ 
by $\Delta f^{\rm equil}_{\rm NL}= {\cal O}(10)$  for an experiment like {\it Planck}. 

\vspace{0.5cm}  \hrule
\def\thefootnote{\arabic{footnote}}
\setcounter{footnote}{0} \vspace{0.5cm}

\section{Introduction}
Cosmological inflation 
\cite{lrreview} has become the dominant paradigm to 
understand the initial conditions for the Cosmic Microwave Background (CMB) anisotropies
and structure formation. 
This picture has recently received further   spectacular confirmation 
by the  Wilkinson Microwave 
Anisotropy Probe (WMAP) five year set of data \cite{wmap5}.
Present \cite{wmap5} and future \cite{planck} experiments 
may be sensitive to the non-linearities of the cosmological
perturbations at the level of second- or higher-order perturbation theory.
The detection of these non-linearities through the  non-Gaussianity
(NG) in the CMB \cite{review} has become one of the primary experimental targets. 

A possible source of NG could be primordial in 
origin, being specific to a particular mechanism for the generation of the cosmological perturbations. This is what 
makes a positive detection of NG so
relevant: it might help in discriminating among competing scenarios which otherwise might be undistinguishable. Indeed,
various models of inflation, firmly rooted in modern 
particle physics theory,   predict a significant amount of primordial
NG generated either during or immediately after inflation when the
comoving curvature perturbation becomes constant on super-horizon scales
\cite{review}. While single-field  \cite{noi}
and two(multi)-field \cite{two} models of inflation generically predict a tiny level of NG, 
`curvaton-type models',  in which
a  significant contribution to the curvature perturbation is generated after
the end of slow-roll inflation by the perturbation in a field which has
a negligible effect on inflation, may predict a high level of NG \cite{luw,ngcurv}.
Alternatives to the curvaton model are those models 
characterized by the curvature perturbation being 
generated by an inhomogeneity in 
the decay rate \cite{hk,varcoupling} or 
  the mass   \cite{varmass} or 
of the particles responsible for the reheating after inflation. 
Other opportunities for generating the curvature perturbation occur
 at the end of inflation \cite{endinflation} and  during
preheating \cite{preheating}.
All these models generate a level of NG which is local as the NG part of the primordial curvature
perturbation is a local function of the Gaussian part, being generated on superhorizon scales. 
In momentum space, the three point function, or bispectrum, arising from the local NG is dominated by the
so-called ``squeezed'' configuration, where one of the momenta is much smaller than the other two and it is parametrized by
the non-linearity parameter $f_{\rm NL}^{\rm loc}$. Other models, such as DBI inflation
\cite{DBI} and ghost inflation \cite{ghost}, predict a different kind of primordial
NG, called ``equilateral'', because the three-point function for this kind of NG is peaked on equilateral configurations, 
in which the lenghts 
of the three wavevectors forming a triangle in Fourier
space are equal \cite{Shapes}. The equilateral NG is parametrized by an amplitude $f_{\rm NL}^{\rm equil}$~\cite{CN}. Present limits
on NG are summarized by $-9<f^{\rm loc}_{\rm NL}<111$ and   $-151<f^{\rm equil}_{\rm NL}<253$ at 95\% CL \cite{wmap5,Curto}.

On the other hand there exist many sources of NG in the CMB anisotropies beyond the primordial ones, which are essential 
to characterize in order to distinguish them from a possible primordial signal. 
One should account for the  so-called  secondary anisotropies,
which arise  after the last scattering epoch. 
For example, 
cross-correlations SZ-lensing and ISW-lensing~\cite{GoldSperg,komatsuspergel} produce a bias in the
estimate of NG which is at the level of the expected estimator
variance at Planck angular resolution \cite{SerraCooray}. Analogous conclusions have been reached
in Ref.~\cite{BabichPierpaoli} for the
cross correlations of density and lensing magnification of radio and SZ
point sources with the ISW effect. Furthermore, the impact of cosmological parameters' uncertainties on estimates of the primordial NG 
parameter in local and equilateral models of NG has been recently studied in \cite{liguoririotto}. 

There exists  another relevant source of NG: the non-linear effects operating at
the recombination epoch. The dynamics at recombination is quite involved because
all the non-linearities in the evolution of the baryon-photon fluid at recombination and  the ones 
coming from general relativity should be accounted for. The first steps in describing the physics at recombination at second-order
in perturbation theory were taken in \cite{CMB2I,CMB2II} (see also Ref.~\cite{Crec,fr}), where 
the full system of Boltzmann equations at second-order describing the evolution of the photon, baryon and cold dark matter fluids were obtained (see also 
Ref.~\cite{P})
These equations allow to follow the time evolution of the CMB anisotropies at second-order at all angular 
scales from the early epoch, when the cosmological perturbations were generated, to the present through the recombination era.
Such a contribution is so relevant because it 
represents a major part of the second-order radiation transfer function which must be determined in order to have a 
complete control of both the primordial and non-primordial part of NG in the CMB anisotropies and to gain from the theoretical side 
the same level of precision that could be reached experimentally in the near future~\cite{review}.

The NG generated at the surface of last scattering comprises various effects, as described in details in Ref.~\cite{CMB2II} 
(for some specific effects see also~\cite{Crec,Wand}). 
It turns out that the dominant contribution comes from the non-linear evolution of the second-order gravitational potential  
which grows in time on small scales. Since this effect is a causal one, developing on small scales, we expect that the NG it generates 
will be of the equilateral type, rather than of local type. As we will see our results confirm such an expectation. 
Therefore, a reasonable
question is to which extent the NG from recombination alters the possibile detection of the primordial NG of the equilateral type.
The goal of this paper is to  estimate in a semi-analytical way the contribution to NG from recombination. Along the same lines of 
Ref.~\cite{BZ},
we adopt a simple analytical  model to parametrize the transfer functions and test its goodness in Section 2 by evaluating the minimum value
of equilateral NG detectable by the Planck experiment for which  there exist numerical calculations \cite{SZ}. 
As a by-product of our 
results, we find that the signal-to-noise ratio for a primordial equilateral bispectrum scales as the 
square root of the maximum multipole $\ell_{\rm max}$ 
probed by an experiment, unlike the well known scaling as $\ell_{\rm max}$ for the local case. 
In Section 3 we compute the bispectrum of the NG generated by the evolution of the second-order gravitational potentials, and in 
Section 4 we provide an estimate
of the NG from recombination showing that it corresponds to a degradation in the 
measurement of an equilateral primordial bispectrum 
of $\Delta f^{\rm equil}_{\rm NL}= {\cal O}(10)$, shifting the minimum detectable value from $f^{\rm equil}_{\rm NL}\simeq 67$ to 
$f^{\rm equil}_{\rm NL}\simeq 79$ for an experiment like {\it Planck}.

\section{Signal-to-Noise ratio for the primordial equilateral bispectrum}
In this Section we wish to recover the estimate for the signal-to-noise ratio $(S/N)$ given in  Ref.~\cite{SZ} for the 
primordial bispectra of ``equilateral'' type~\cite{CN} by adopting a simple model. In other words,  we test the goodness
of the semi-analytical model we will be using in the next Section to   estimate the bispectrum from the recombination era.

Our starting point is the primordial equilateral bispectrum~\cite{CN}
\be
\label{equil}
\langle \Phi({\bf k}_1) \Phi({\bf k}_2) \Phi({\bf k}_3) \rangle = (2 \pi)^3 \delta^{(3)}
\big({\bf k}_1 + {\bf k}_2 + {\bf k}_3 \big)
B_{\rm equil}( k_1,  k_2 ,  k_3) \, ,
\ee
where
\be
\label{eq:ours}
B_{\rm equil}(k_1,k_2,k_3) = f_{\rm NL}^{\rm equil} \cdot 6  
 A^2 \cdot \left(-\frac1{k_1^3 k_2^3} - 
\frac1{k_1^3 k_3^3} - \frac1{k_2^3 k_3^3}  - \frac2{k_1^2 k_2^2 k_3^2} + \frac1{k_1 k_2^2 k_3^3}
+ (5 \; {\rm perm.}) \right)\, , 
\ee
and  the permutations act only on the last term in parentheses. The parameter $f_{\rm NL}^{\rm equil}$ quantifies the level of NG while
$A=17.46 \times 10^{-9}$ 
is the amplitude of the  primordial gravitional potential 
power spectrum computed at first-order

\be
\langle \Phi^{(1)}({\bf k}_1) \Phi^{(1)}({\bf k}_2) \rangle = (2 \pi)^3 \delta^{(3)}
\big({\bf k}_1 + {\bf k}_2 \big) P(k_1)\, ,
\ee 
with $P(k)=A/k^3$. Since the signal-to-noise ratio $(S/N)$ will be some function of the maximum multipole a given 
experiment can reach, $\ell_{\rm max}\gg 1$, we can use the flat-sky approximation ~\cite{BZ,Hulensing} and write
for the bispectrum 

\be
\langle a(\vec{\ell}_1)a(\vec{\ell}_2)a(\vec{\ell}_3) \rangle 
   = (2\pi)^2\delta^{(2)}(\vec{\ell}_{123}) B(\ell_1,\ell_2,\ell_3)\, ,
\ee 
where $\vec{\ell}_{123}=\vec{\ell}_1+\vec{\ell}_2+\vec{\ell}_3$, 
with~\cite{Shapes}  
\begin{equation} 
\label{bispectflat}
B_{\rm equil}(\ell_1,\ell_2,\ell_3) = \frac{(\tau_0 - \tau_r)^2}{(2\pi)^2} \int
dk^z_1dk^z_2dk^z_3 \delta^{(1)}(k^z_{123}) B_{\rm equil}(k'_1,k'_2,k'_3)  
\tilde{\Delta}^T(\ell_1,k^z_1) \tilde{\Delta}^T(\ell_2,k^z_2) \tilde{\Delta}^T(\ell_3,k^z_3)\, ,
\end{equation}
where $k'$ means $k$ evaluated such that $\vec{k}^{\parallel} = \vec{\ell}/(\tau_0-\tau_r)$ and 
   \begin{equation}
        \tilde{\Delta}^T(\ell,k^z) = \int_0^{\tau_0} \frac{d\tau}{(\tau_0-\tau)^2} 
          S(\sqrt{(k^z)^2 + \ell^2/(\tau_0-\tau)^2},\tau) 
          e^{ik^z(\tau_r-\tau)} \, , 
   \end{equation}
is the radiation transfer function defined by the CMB source function $S(k,\tau)$. In this notation, $\tau_0$ and $\tau_r$  represent the
present-day and the recombination conformal time, respectively and $k^z$ and $\vec{k}^{\parallel}$ are the momentum components
perpendicular and parallel respectively to the plane orthogonal to the line-of-sight.

The $(S/N)$ ratio in the flat-sky formalism is~\cite{BZ,Hulensing} 

\begin{equation}
\label{S/N}
\left( \frac{S}{N} \right)^2= 
\frac{f_{\rm sky}}{\pi} \frac{1}{(2 \pi)^2} \int d^2 \ell_1 d^2 \ell_2  d^2 \ell_3 
\,\delta^{(2)}(\vec{\ell}_{123})\, 
\frac{B_{\rm equil}^2(\ell_1,\ell_2,\ell_3)}{6\, C(\ell_1)\,C(\ell_2)\, C(\ell_3)}\, ,
\end{equation}
where $f_{\rm sky}$ stands for  the portion of the observed sky. 
In order to compute the bispectrum $B_{\rm equil}(\ell_1,\ell_2,\ell_3)$ 
and the power spectrum $C(\ell)$ we adopt the following model   
\be
\label{toymodel}
a({\vec \ell})= \int \frac{d k^z}{2\pi} e^{ik^z(\tau_0-\tau_r)} \Phi({\bf k}') \tilde{\Delta}^T(\ell,k^z)
\ee 
where we mimic the effects of the transfer function on small scales as 

\be
\tilde{\Delta}^T(\ell,k^z)=a\,(\tau_0-\tau_r)^{-2}  e^{-1/2(\ell/\ell_*)^{1.2}}
e^{-1/2(|k_z|/ k_*)^{1.2}}\, ,
\ee 
{\it i.e.} a simple exponential and a normalization 
coefficent $a$ to be 
determined to match the amplitude of the angular power spectrum at the characteristic scale 
$\ell \simeq \ell_*=k_* (\tau_0-\tau_r)$.\footnote{We could equally choose a transfer function as 
$\tilde{\Delta}^T(\ell,k^z)=a\, (\tau_0-\tau_r)^{-2}\,e^{-1/2(\ell/\ell_*)^{1.2}}  \theta(k_*-|k^z|)$, the relevant approximation 
being that the integral over $k^z$ is cut at the 
scale $k_*$.} It is important to make clear what are the reasons underlying the choice of such a
model. When computing the $(S/N)$, Eq.~(\ref{toymodel}) with $\ell_*=k_*(\tau_0-\tau_r)\simeq750$ and $a\simeq 3$ 
is able to account for the combined effects of ``radiation driving'', 
which occours at $\ell>\ell_{\rm eq}\simeq 160$ and boosts the angular power spectrum with respect to the Sachs-Wolfe plateau, 
and the effects of 
Silk damping which tend to suppress the CMB anisotropies for scales $\ell>\ell_D\simeq 1300$. The combination of these 
effects produces a decrease in the angular power spectrum from a scale $\ell_*\simeq 750$.\footnote{
The choice of the exponent $1.2$ derives from the study of the diffusion damping envelope in Ref. ~\cite{HuWhite}.}  
The power spectrum in 
the flat-sky approximation is given 
by  $\langle a(\vec{l}_1)a(\vec{l}_2) \rangle 
   = (2\pi)^2\delta^{(2)}(\vec{l}_{12}) C(\ell_1)$ with 
\be
C(\ell)=\frac{(\tau_0-\tau_r)^2}{(2 \pi)} \int dk^z |\tilde{\Delta}^T(\ell,k^z)|^2\, P(k)\, .
\ee       
The exponential of the transfer function for Eq.~(\ref{toymodel}) allows to cut off the integral for $k\simeq k_*$ and one finds 
(see also Ref.~\cite{BZ})
\be
\label{Cl}
C(\ell)=a^2 \frac{A}{\pi \ell^2} \frac{e^{-(\ell/\ell_*)^{1.2}}}{\sqrt{1+\ell^2/\ell_*^2}}\simeq a^2 \frac{A}{\pi}\frac{\ell_*}{\ell^3}\,
e^{-(\ell/\ell_*)^{1.2}}\, ,
\ee
where the last equality holds for $\ell\gg\ell_*$. 
To compute the bispectrum we proceed in a similar way. One first uses the Dirac deltas, 
$\delta^{(1)}(k^z_{123})$ and $\delta^{(2)}(\vec{\ell}_{123})$. Then it 
proves to be useful the change of 
variable $k^z_1=x_1 \ell_1/(\tau_0-\tau_r)$, $k^z_2=x_2 \ell_2/(\tau_0-\tau_r)$. 
In this way the transfer functions become $ \tilde{\Delta}^T(\ell_i,k_i^z)\propto 
e^{-1/2(|x_i| \ell_i/\ell_*)^{1.2}}$ which allows to cut the integrals over $x_i$ ($i=1,2$) at $\ell_*/\ell_i$. Now, as a good approximation to see the effects of the transfer functions, we can take $\ell \gg \ell_*$ and thus the integral over $x_i$ can be easily computed by just evaluating the integrand 
in $x_i=0$ times $4(\ell_*/\ell_1)(\ell_*/\ell_2) $. 
With this approximation the integral in $k^z_i$ is easily obtained and we get for the bispectrum 
\begin{eqnarray}
\label{Bapprox}
B_{\rm equil}(\ell_1,\ell_2,\ell_3)&=& \frac{24 f_1}{(2 \pi)^2}\, f^{\rm equil}_{\rm NL} a^3\, A^2\,  
e^{-(\ell_1^{1.2}+\ell_2^{1.2}+\ell_3^{1.2})/2\ell^{1.2}_*}\,
\ell_*^2 \nonumber\\
&\times&  \left(-\frac1{\ell_1^3 \ell_2^3} - 
\frac1{\ell_1^3 \ell_3^3} - \frac1{\ell_2^3 \ell_3^3}  - \frac2{\ell_1^2 \ell_2^2 \ell_3^2} + \frac1{\ell_1 \ell_2^2 \ell_3^3}
+ (5 \; {\rm perm.}) \right)\, , 
\end{eqnarray}
where 
\be
\label{l3}
\ell_3^2=\ell_1^2+\ell_2^2+2\,\vec{\ell}_1\cdot\vec{\ell}_2\, . 
\ee
The coefficient $f_1\simeq 1/1.4=0.7$ is a fudge factor that improves the matching between our approximation for the bispectrum 
and numerical results that have been consistenly checked. Notice that, according to our approximation, the equilateral structure of 
Eq.~(\ref{eq:ours}) is preserved in $\ell$ space.\footnote{The expression~(\ref{Bapprox}) can be also written as 
\newline
$B_{\rm equil}(\ell_1,\ell_2,\ell_3)=  (2 \pi)^{-2} 48 f_1\, f^{\rm equil}_{\rm NL} \,
a^3\,A^2\,  \ell_*^2 e^{-(\ell_1^{1.2}+\ell_2^{1.2}+\ell_3^{1.2})/2\ell^{1.2}_*}\,(1+\cos\theta) (\ell_1+\ell_2-\ell_3)/
\ell_1^2 \ell_2^2\ell_3^3$, $\theta$ being the angle between ${\vec \ell}_1$ and ${\vec \ell}_2$.}
In computing the signal-to-noise ratio, consistency with our approximation~(\ref{Bapprox}) requires that we integrate over 
$\ell_1,\ell_2$ starting from a minimum $\ell_{\rm min} > \ell_*$ up to $\ell_{\rm max}$ and paying attention to the fact that even $\ell_3$ in 
Eq. (\ref{l3}) must be larger than $\ell_{\rm min}$.  The scaling with $\ell_{\rm max}$ with 
respect to the case of a local type bispectrum turns out to be much milder. 
While for the local type $(S/N)^2 \propto \ell_{\rm max}^2$ \cite{BZ}, 
 for the equilateral 
bispectrum~(\ref{equil}) we find\footnote{
These scalings can be 
easily understood by analyzing the expressions $(S/N)^2$ for the 
local and equilateral primordial NG. In the local case,  $(S/N)^2$ is 
proportional to $\int d^2\ell_1 d^2\ell_2 d^2\ell_3 \delta^{(2)}
(\vec{\ell}_{123})(\ell^3_1+\ell^3_2+\ell^3_3)^2/(\ell_1\ell_2\ell_3)^3$~\cite{BZ};
since the squeezed configuration, {\it e.g.} $\ell_1\ll \ell_2,\ell_3$,  
is dominating the local bispecrum, the integral becomes proportional to $\int 
d\ell_1 d\ell_2 (\ell_2/\ell^2_1)\propto \ell^2_{\rm max}$. In the equilateral
case, however, $(S/N)^2$ receives contributions from the configuration which is peaked at $\ell_1\sim\ell_2\sim\ell_3$ and therefore 
it can be written as  $\int d^2\ell_1
d^2\ell_2 \delta(\ell_1-\ell_2)/\ell_1^2\propto \ell_{\rm max}$.}  
$(S/N)^2 \propto \ell_{\rm max}$ and, setting $\ell_*=750$ and $\ell_{\rm min}\simeq 1200$, 
\begin{equation}
\label{SNprim}
\left( \frac{S}{N} \right)_{\rm equil}^2 = 0.48 \times 10^5\, \frac{f_{\rm sky}}{2^5 \pi^3 6} \,A\, (f^{\rm equil}_{\rm NL})^2\, \, 
\ell_{\rm max} \simeq  
8 \, f_{\rm sky} \,A\,   (f^{\rm equil}_{\rm NL} )^2\,  
\ell_{\rm max} . 
\end{equation}
By choosing $f_{\rm sky}=0.8$ and $\ell_{\rm max}=2000$ we find a minimum detectable 
\be
\label{fnleq}
f^{\rm equil}_{\rm NL}\simeq 66\, ,
\ee
obtained imposing $(S/N)_{\rm equil} =1$. Both the estimate of the minimum value of 
$f^{\rm equil}_{\rm NL}$ and the scaling $(S/N)^2 \propto \ell_{\rm max}$ are 
 in remarkable agreement with the result obtained in Ref.~\cite{SZ}  
where the full transfer function is used and a value of $f^{\rm equil}_{\rm NL}= 67$ is obtained.\footnote{We thank M. Liguori for discussions about 
the minimum value of $f^{\rm equil}_{\rm NL}$ detectable by Planck
and for its scaling with $\ell_{\rm max}$.} Notice that our estimate is independent from the coefficient $a$ and the exponetial $e^{-1/2(\ell/\ell_*)^{1.2}}$ 
introduced below Eq.~(\ref{toymodel}) to mimic the full transfer function.  This is because  there is an equal number of transfer 
functions in the
numerator and denominator of the expression (\ref{S/N}) for the signal-to-noise ratio and their effect tend to cancel despite they 
are not simple
multiplicative factors (see discussion in Ref.~\cite{BZ}).

\section{Non-Gaussianity from recombination}
Comforted by the goodness of our model, in this Section we wish to estimate the level of NG generated at the recombination era. One can 
check that on small scales the second-order anisotropies are dominated 
by the second-order gravitational potential $\Phi^{(2)}$ which grows as $\tau^2$, as first pointed out in Ref.~\cite{fr}.\footnote{
Notice that we have numerically 
verified that such a term starts to be relevant from multipoles $\ell \simeq 1300$. This is actually indicated also by the results of 
Ref.~\cite{fr}.}
Therefore, the  main contribution to the bispectrum generated at recombination comes from   
\begin{eqnarray}
\label{monopole}
\Theta^{(2)}_{\rm SW}=\frac{1}{4} \Delta^{(2)}_{00}+\Phi^{(2)}\, ,
\end{eqnarray}
which is the usual term appearing in the CMB anisotropies due to the intrinsic photon energy density fluctuations 
$\Delta^{(2)}_{00}$ and the gravitational redshift due to the potential~\cite{CMB2II}. Such a term on large scales reduces to the 
Sachs-Wolfe effect, while on small scales at recombination 
\be
\label{dominant}
\frac{1}{4} \Delta^{(2)}_{00}+\Phi^{(2)} \simeq -R\,  \Phi^{(2)}=-\frac{R}{14} G({\bf k_1}, {\bf k_2},{\bf k})T(k_1) \Phi^{(1)}({\bf k}_1)
 T(k_2)\Phi^{(1)}({\bf k}_2)\,\tau^2_r\, ,
\ee
where we have evaluated the expression at the recombination time $\tau_r$ and $R=3\rho_{\rm b}/4\rho_\gamma$ is the
baryon-to-photon energy density ratio. Eq.~(17) is the extension to the second-order of a well-known
expression at linear order~\cite{HuWhite} (see also \cite{HuS,PhDHu}), as discussed in details in Ref.~\cite{fr}. It can also be easily obtained
using the expressions of Ref.~\cite{CMB2II} by taking the sound speed
as a function of $R$. The kernel is 
 given by~\cite{CMB2II} 
\be
\label{kernelG}
G({\bf k_1}, {\bf k_2},{\bf k})={\bf k_1} \cdot {\bf k_2}-\frac{10}{3} \frac{({\bf k}\cdot {\bf k_1})
({\bf k}\cdot {\bf k_2})}{k^2}\, .
\ee  
Notice that this expression has been obtained assuming all the momenta much larger than $k_{\rm eq}$ \cite{CMB2II}. 
From the form of the kernel we see that the NG at recombination is dominated by an equilateral configuration, as expected 
from the fact that its origin is gravitational. 
Here and in the following we are implicitly assuming that a convolution is acting on the kernel as  
\be
\frac{1}{(2 \pi)^3} \int d{\bf k}_1 d{\bf k}_2\, \delta^{(3)}({\bf k}_1+{\bf k}_2+{\bf k}_3) G({\bf k_1}, {\bf k_2},{\bf k})  
T(k_1)\Phi^{(1)}({\bf k}_1)
T(k_2) \Phi^{(1)}({\bf k}_2)\, .
\ee
The reader should remember that, at first-order in perturbation theory,  the  combination
$\Theta^{(1)}_{\rm SW}+R\,\Phi^{(1)}$ is exponentially suppressed by the Silk damping, but still greater than the 
term $R\Phi^{(1)}$ (which does not suffer the damping) for the maximum multipole of interest, $\ell_{\rm max}\sim 2000$. This is meanly due to the fact that the first-order gravitational potential rapidly decays on small scales. On the contrary, at second-order in perturbation theory, the
gravitational potential grows like the scale factor on small scales and it turns out that the $R\Phi^{(2)}$ dominates on small scales 
(see Ref.~\cite{fr}).

The gravitational potential at linear order can be expressed as usual in terms of the transfer function  $T(k)$
\be
T(k) \approx 12 \left( \frac{k_{\rm eq}}{k} \right)^2 \ln[k/8 k_{\rm eq}]\, ,  
\ee
where the last step is an approximation valid on  scales smaller the the equivalence scale, $k\gg k_{\rm eq}$.
In the following we will account for the logarithmic growth just with a 
coeffcient $T_0(k)=12 \ln[k/8 k_{\rm eq}] \approx 11$ for the scales of interest. 

In the flat-sky approximation one arrives at an expression similar to~(\ref{bispectflat}), where now one of the linear 
transfer functions must replaced by a transfer function at second-order. Specifically one finds
\begin{eqnarray} 
\label{bis}
B_{\rm rec}(\ell_1,\ell_2,\ell_3)&=& \frac{(\tau_0 - \tau_r)^2}{(2\pi)^2} \int
dk^z_1dk^z_2dk^z_3 \delta^{(1)}(k^z_{123}) \Big[ G({\bf k}'_1,{\bf k}'_2,{\bf k}'_3)T(k'_1) T(k'_2)P(k'_1) P(k'_2)  
\\ \nonumber
& \times &\tilde{\Delta}^{T}(\ell_1,k^z_1) \tilde{\Delta}^T(\ell_2,k^z_2) \tilde{\Delta}^{T(2)}(\ell_3,k^z_3) +
{\rm cyclic} \Big]\, .
\end{eqnarray}
By using our model~(\ref{toymodel}) and 
\be
\tilde{\Delta}^{T(2)}(\ell,k^z)=- \frac{R}{14} \frac{\tau^2_r}{(\tau_0-\tau_r)^2}\, ,   
\ee
for the second-order radiation transfer function, we find
\bea
B_{\rm rec}(\ell_1,\ell_2,\ell_3)&=&-\frac{R}{14} \frac{(\tau_0-\tau_r)^{-4}}{(2 \pi)^2}k^4_{\rm eq}\, \tau_r^2A^2 a^2 T_0^2 
e^{-1/2(\ell_1/\ell_*)^{1.2}} e^{-1/2(\ell_2/\ell_*)^{1.2}}  \int
dk^z_1dk^z_2dk^z_3 \delta^{(1)}(k^z_{123}) \nonumber \\
&\times& \left[ G({\bf k}'_1,{\bf k}'_2,{\bf k}'_3) \frac{1}{{k'_1}^5 {k'_2}^5}  e^{-1/2(|k_{1z}|/ k_*)^{1.2}} e^{-1/2(|k_{2z}|/ k_*)^{1.2}} +{\rm cyclic} \right]\, . 
\eea
At this point we proceed further by employing the same approximation described after Eq. (\ref{Cl}). We 
use the Dirac delta to replace the variable $k_{3z}$, and the exponential allow us to 
evaluate the integral for $k_{1z}=k_{2z}=0$, for scales $\ell_i \gg \ell_*$. This leads to 
\bea
\label{Bisprec}
B_{\rm rec}(\ell_1,\ell_2,\ell_3)&=&-\frac{4 f_2}{(2 \pi)^2}\frac{R}{14} A^2 a^2 T_0^2 (k_{\rm eq} \tau_r)^2 \ell_{\rm eq}^2 \ell_*^2 
e^{-1/2(\ell_1/\ell_*)^{1.2}} e^{-1/2(\ell_2/\ell_*)^{1.2}}\nonumber\\
&\times& \frac{1}{\ell_1^5 \ell_2^5} \Big[{\vec\ell}_1 \cdot {\vec\ell}_2-\frac{10}{3} 
\frac{({\vec\ell}_3 \cdot {\vec\ell}_1)({\vec\ell}_3 \cdot {\vec\ell}_2)}{\ell_3^2} \Big]+ 
 {\rm cyclic}\, .
\eea
Again here $f_2$ is a coefficient to better calibrate our approximations with numerical results that we have performed in order to 
test the validity of our approach. Not surprisingly, it turns out that $f_2\simeq f_1 \simeq 1/1.4$.  
\section{Contamination to primordial non-Gaussianity from recombination: Fisher matrices}
Our goal now is to quantify the level of NG coming from the recombination era and to estimate the level of degradation it causes
on the possible measurement of the equilateral primordial bispectrum. The reader should keep in mind that, given the form 
of the kernel function (\ref{kernelG}), the NG from recombination is expected to be of the equilateral type. 
A rigorous procedure is to define the Fisher matrix (see, for example, \cite{komatsuspergel})
\be 
F_{ij}=\int d^2 \ell_1 d^2 \ell_2  d^2 \ell_3 
\,\delta^{(2)}(\vec{\ell}_{123})\,\frac{
B^{i}(\ell_1,\ell_2,\ell_3)\, B^{j}(\ell_1,\ell_2,\ell_3)}{6\, C(\ell_1)\,C(\ell_2)\, C(\ell_3)}\, ,
\ee
where $i$ (or $j$)$=({\rm rec},{\rm equil})$, and to define the signal-to-noise ratio for a component $i$, 
$(S/N)_i=1/\sqrt{F^{-1}_{ii}}$,  and the degradation parameter $d_i=F_{ii} F^{-1}_{ii}$ due to the correlation bewteen the 
different components $r_{ij}=F^{-1}_{ij}/\sqrt{F^{-1}_{ii}F^{-1}_{jj}}$. The first entry $F_{\rm equil,equil}$ 
of the Fisher matrix corresponds to 
the $(S/N)^2$ ratio computed in Eq.~(\ref{SNprim}) which does not account for any kind of cross-correlation. 
Due to the equilateral form of the 
NG generated at recombination we expect that the minimum value detectable for $f_{\rm NL}^{\rm equil}$ will 
be higher that the one reported in Eq.~(\ref{fnleq}). For the mixed entry we find

\bea
\label{S/Nij}
F_{\rm rec,equil}&=& 
\frac{f_{\rm sky}}{\pi} \frac{1}{(2 \pi)^2} \int d^2 \ell_1 d^2 \ell_2  d^2 \ell_3 
\,\delta^{(2)}(\vec{\ell}_{123})\, 
\frac{B_{\rm rec}(\ell_1,\ell_2,\ell_3)  B_{\rm equil}(\ell_1,\ell_2,\ell_3)}
{6\, C(\ell_1)\,C(\ell_2)\, C(\ell_3)}\, \nonumber\\
&=&- 3 f_1 f_2 \frac{f_{\rm sky}}{\pi^3} \frac{4 R}{14} \frac{48}{2^5 6}\frac{T^2_0}{a} 
\,(k_{\rm eq} \tau_r)^2 \ell^2_{\rm eq} \ell_*\, A f_{\rm NL}^{\rm equil}
\, \int d\ell_1 d\ell_2 (1+\vec{\ell}_1\cdot\vec{\ell}_2/\ell_1\ell_2)\, 
e^{1/2(\ell_3/\ell_*)^{1.2}} \nonumber\\
&\times& \frac{1}{\ell_1^3 \ell_2^3}\left(\ell_1+\ell_2 -\ell_3 \right) \left[{\vec\ell}_1 \cdot {\vec\ell}_2-\frac{10}{3} 
\frac{({\vec\ell}_3 \cdot {\vec\ell}_1)({\vec\ell}_3 \cdot {\vec\ell}_2)}{\ell_3^2} \right]\, ,
\eea 
where $\ell_3$ is given by Eq.~(\ref{l3}). The factor 3 in front of this expression comes from cyclic permutations. 
The integral can be performed numerically and, 
integrating from a minimum $\ell_{\rm min}\simeq 1200$ up to $\ell_{\rm max}=2000$, and 
by taking $R\simeq 0.3$ when evaluated at recombination, $a \simeq 3$, $T_0 \simeq 11$, 
$(k_{\rm eq} \tau_r)^2\simeq 26$, $\ell_{\rm eq}=150$, $\ell_*=750$, we find $F_{\rm rec,equil} \simeq 
9.4 \times 10^{-4}$. Finally for the entry $F_{\rm rec,rec}$ we get

\bea
\label{S}
F_{\rm rec,rec}&=&  
\frac{f_{\rm sky}}{\pi} \frac{1}{(2 \pi)^2} \int d^2 \ell_1 d^2 \ell_2  d^2 \ell_3 
\,\delta^{(2)}(\vec{\ell}_{123})\, 
\frac{B^2_{\rm rec}(\ell_1,\ell_2,\ell_3)}
{6\, C(\ell_1)\,C(\ell_2)\, C(\ell_3)}\, \nonumber\\
&=& f_2^2 \frac{f_{\rm sky}}{2^5 \pi^3 6} \left( \frac{4 R}{14} \right)^2 \left( \frac{T^2_0}{a} \right)^2 
(k_{\rm eq} \tau_r)^4\, \ell^4_{\rm eq} \ell_*  A\Bigg[
3 \int d\ell_1 d\ell_2 e^{(\ell_3/\ell_*)^{1.2}} \frac{\ell_3^3}{\ell_1^6 \ell_2^6} \left({\vec\ell}_1 \cdot {\vec\ell}_2-\frac{10}{3} 
\frac{({\vec\ell}_3 \cdot {\vec\ell}_1)({\vec\ell}_3 \cdot {\vec\ell}_2)}{\ell_3^2} \right)^2\nonumber \\   
&+& 6 \int d\ell_1 d\ell_2  e^{1/2(\ell_3/\ell_*)^{1.2}}  e^{1/2(\ell_2/\ell_*)^{1.2}} \frac{1}{\ell_1^6 \ell_2 \ell_3^2} 
 \left({\vec\ell}_1 \cdot {\vec\ell}_2-\frac{10}{3} 
\frac{({\vec\ell}_3 \cdot {\vec\ell}_1)({\vec\ell}_3 \cdot {\vec\ell}_2)}{\ell_3^2} \right)\nonumber \\
&\times& \left({\vec\ell}_1 \cdot {\vec\ell}_3-\frac{10}{3} 
\frac{({\vec\ell}_2 \cdot {\vec\ell}_3)({\vec\ell}_1 \cdot {\vec\ell}_2)}{\ell_2^2} \right)
\Bigg]\, , 
\eea
and we find a value $F_{\rm rec, rec}\simeq 0.014$. We are now able to compute the entries of inverse of the Fisher matrix, $F^{-1}_{ij}$. In 
the following we report our results for the signal-to-noise ratios and the degradation parameters
\bea
\label{SN1}
\left( \frac{S}{N} \right)_{\rm equil}&=& \frac{1}{\sqrt{F^{-1}_{\rm equil,equil}}} \simeq
12.6 \times 10^{-3}  f_{\rm NL}^{\rm equil}
\, ,\\
\left( \frac{S}{N} \right)_{\rm rec}&=& \frac{1}{\sqrt{F^{-1}_{\rm rec,rec}}}\simeq 0.1\, , \\ 
r_{\rm rec,equil}&=&\frac{F^{-1}_{\rm rec,equil}}{\sqrt{F^{-1}_{\rm equil,equil}F^{-1}_{\rm rec,rec}}}\simeq -0.53\, , \\
d_{\rm rec}&=&F_{\rm rec,rec}F^{-1}_{\rm rec,rec}\simeq 1.4\, , \\
d_{\rm equil}&=&F_{\rm equil,equil}F^{-1}_{\rm equil,equil}\simeq 1.4\, .
\eea
As a confirmation of our expectations, we find that the NG of the type given by Eq.~(\ref{kernelG}) has a quite high correlation with an 
equilateral primordial bispectrum.
This translates into a degradation (or a contamination) in the mimimum detectable value for 
$f_{\rm NL}^{\rm equil}$ with respect to the value given in~(\ref{fnleq}). In fact from the signal-to-noise ratio~(\ref{SN1}) we find 
a minimum value of 
\be
f_{\rm NL}^{\rm equil} \simeq 79\, ,
\ee  
imposing that $(S/N)_{\rm equil}=1$. This roughly corresponds to a contamination to the primordial equilateral NG 
of\footnote{Due to a non-vanishing correlation, 
$r_{ij}$, $(S/N)$ gets modified from its zero-order value to $(S/N)=(S/N)_0 (1-r_{ij}^2)^{1/2}$, so that the minimum detectable value of $f^{\rm equil}_{\rm NL}$ gets shifted by a quantity $\Delta f^{\rm equil}_{\rm NL}/(f^{\rm equil}_{\rm NL})_0 =(1-r_{ij}^2)^{-1/2}-1$.} 

\be
\label{kk}
\Delta f^{\rm equil}_{\rm NL}={\cal O}(10)\, .
\ee
A similar way to quantify this statement is by defining an ``effective'' $f^{\rm rec}_{\rm NL}$ for which the equilateral 
bispectrum~(\ref{eq:ours}) has the same Fisher matrix errors as the recombination bispectrum (see also ~\cite{Liguorietal,SZ})
\be
\label{eff}
f^{\rm rec}_{\rm NL}=\frac{\sqrt{F_{\rm rec,rec}}}{\sqrt{F_{\rm equil,equil}}}\Big|_{f^{\rm equil}_{\rm NL}=1}\, .
\ee
In this case we find an effective  non-linearity parameter $f^{\rm rec}_{\rm NL} \simeq 8$ which agrees with the result
(\ref{kk}).\footnote{
An alternative quantity can be used in order to measure the contamination to the primordial bispectra. 
It is that effective equilateral  $f^{\rm rec}_{\rm NL}$ which minimizes the $\chi^2$ defined as 
\be
\chi^2=\int d^2 \ell_1 d^2 \ell_2  d^2 \ell_3 
\,\delta^{(2)}(\vec{\ell}_{123})\,
\frac{\left(\left. f^{\rm rec}_{\rm NL}\,B_{\rm eq}(\ell_1,\ell_2,\ell_3)\right|_{f_{\rm NL}^{\rm equil}=1}-
B_{\rm rec}(\ell_1,\ell_2,\ell_3)\right)^2}{6\, C(\ell_1)\,C(\ell_2)\, C(\ell_3)}\, . \nonumber 
\ee
One finds 
\be
\label{effr}
f^{\rm rec}_{\rm NL}= \frac{F_{\rm rec,equil}}{F_{\rm equil ,equil}} \Big|_{f_{\rm NL}^{\rm equil}=1}\, , \nonumber
\ee
and an analogous expression to compute the contamination to the local primordial bispectrum. In both cases we find a similar 
value to the one obtained from Eq.~(\ref{eff})
and Eq.~(\ref{effloc}). Notice however that the effective non-linearity parameter defined in this way contains  
a somewhat richer information with respect to~(\ref{eff}): we are not just comparing signal-to-noise ratios, but we are asking what is 
the value of equilateral (local) $f_{\rm NL}$ which best mimics the bispectrum from recombination.    
} 
Similarly we can compute the Fisher matrix accounting for the NG generated at recombination and the primordial NG of the local type 
\be
\label{equilloc}
\langle \Phi({\bf k}_1) \Phi({\bf k}_2) \Phi({\bf k}_3) \rangle = (2 \pi)^3 \delta^{(3)}
\big({\bf k}_1 + {\bf k}_2 + {\bf k}_3 \big)
B_{\rm loc}( k_1,  k_2 ,  k_3) \, ,
\ee
where
\be
\label{eq:loc}
B_{\rm loc}(k_1,k_2,k_3) = f_{\rm NL}^{\rm loc} \cdot 2  
 A^2 \cdot \left(\frac1{k_1^3 k_2^3} +
\frac1{k_1^3 k_3^3} + \frac1{k_2^3 k_3^3}  \right) \, .
\ee
The bispectrum and the signal-to-noise ratio as defined in Eq.~(\ref{S/N}) have already been computed in the flat-sky approximation in 
Ref.~\cite{BZ}. The result is that $(S/N)_{\rm loc}^2=4 \pi^{-2} 
 f_{\rm sky} (\ell_*/\ell_{\rm min}) (f^{\rm loc}_{\rm NL})^2 
A\, \ell^2_{\rm max}$, corresponding to 
a minimum 
detectable value of $f^{\rm loc}_{\rm NL}={\cal O}(7)$ for $\ell_{\rm max}=2000$ 
(when other possible sources of NG are ignored). We can compute the 
off-diagonal entry of the Fisher matrix in a similar way to what we have described in this section, and we get 
$F_{\rm rec,loc}\simeq 8 \times 10^{-3} f_{\rm NL}^{\rm loc}$. Finally the entry $F_{\rm rec,rec}\simeq 0.014$ 
has already been computed above. 
From inverting the Fisher matrix, we get the following signal-to-noise ratios and the degradation parameters
\bea
\label{SN2}
\left( \frac{S}{N} \right)_{\rm loc}&=& \frac{1}{\sqrt{F^{-1}_{\rm loc,loc}}} \simeq 14 \times 10^{-2}  f_{\rm NL}^{\rm loc}
\, ,\\
\left( \frac{S}{N} \right)_{\rm rec}&=& \frac{1}{\sqrt{F^{-1}_{\rm rec,erec}}}\simeq 0.1\, , \\ 
r_{\rm rec,loc}&=&\frac{F^{-1}_{\rm rec,loc}}{\sqrt{F^{-1}_{\rm loc,loc}F^{-1}_{\rm rec,rec}}}\simeq -0.44\, , \\
d_{\rm rec}&=&F_{\rm rec,rec}F^{-1}_{\rm rec,rec}\simeq 1.2\, , \\
d_{\rm loc}&=&F_{\rm equil,equil}F^{-1}_{\rm equil,equil}\simeq 1.2\, .
\eea
In particular,  from Eq.~(\ref{SN2}) we see that now the minimum detectable value of $f^{\rm loc}_{\rm NL}$ remains basically
 unchanged in the presence of the recombination signal. 
Similarly the effective $f^{\rm rec}_{\rm NL}$ reads
\be
\label{effloc}
f^{\rm rec}_{\rm NL}=\frac{\sqrt{F_{\rm rec,rec}}}{\sqrt{F_{\rm loc,loc}}}\Big|_{f^{\rm loc}_{\rm NL}=1} \simeq 0.7 \, ,
\ee
which is much smaller than the effective non-linearity parameter~(\ref{eff}) for the equilateral case. 
We have also checked the cross-correlation between 
the primordial local and equilateral bispectra finding a value of $r_{\rm loc,equil}\simeq0.23$, which is in 
agreement with the 
value reported in Ref.~\cite{SZ}. This reflects the fact that the primordial local and equilateral signals are 
not fully uncorrelated. The reason  is due to the fact that the equilateral and 
local bispectrum~(\ref{eq:loc}) and~(\ref{eq:ours}) 
approach the same shape in the equilateral configuration. This is also the reason why the 
cross-correlation between the primordial local and recombination bispectra is not so small.

We conclude with two comments. First, we would like to stress that 
the NG from recombination and the one due to the non-linear evolution 
of gravity from the last scattering surface to us include many other contributions.  However, they will affect 
the measurement of the primordial local NG \cite{Nitta}, while the contamination to the primordial equilateral NG is dominated
by the non-linearities considered in this note. 
Second, our estimates have been obtained  with a  maximum multipole of $\ell_{\rm max}=2000$. For an experiment 
like {\it Planck}, it  approximately corresponds  to the angular scales where the instrument noise and the 
secondary effects from lensing are still negligble 
when computing the signal-to-noise ratios~\cite{komatsuspergel,BZ}.
           
\vskip 0.5cm
\begin{center}
{\bf Acknowledgments}
\end{center}
We thank M. Liguori, E. Komatsu and S. Matarrese for several discussions. This research has been partially supported by ASI contract 
I/016/07/0 ``COFIS''.

\end{document}